\documentclass[runningheads]{llncs}
\usepackage[utf8]{inputenc}
\usepackage{graphicx}
\usepackage{url}
\usepackage{caption}

\usepackage{multirow}
\usepackage{booktabs}
\usepackage{longtable}
\usepackage{booktabs}

\usepackage{hyperref}
\usepackage{comment}
\usepackage{subcaption}
\usepackage[
style=nature,
isbn=false,
citestyle=numeric,
natbib=false,
doi=false,
url=false,
maxcitenames=2,
urldate=iso,
date=iso,
seconds=false
]{biblatex}
\newcommand{\FIPER}{\emph{FIPER}}
\begin{document}

\title{FIPER: a Visual-based Explanation Combining Rules and Feature Importance}
%
%
\author{Eleonora Cappuccio \inst{1,2,3}\orcidID{0000-0002-6105-2512} \and
Daniele Fadda \inst{2}\orcidID{0000-0002-0051-0604} \and
Rosa Lanzilotti \inst{3}\orcidID{0000-0002-2039-8162} \and
Salvatore Rinzivillo\inst{2}\orcidID{0000-0003-4404-4147}}
\authorrunning{E. Cappuccio et al.}
%
\institute{Università di Pisa, Italy\\ \email{\{name.surname\}@phd.unipi.it} \and
Università degli Studi di Bari Aldo Moro\\
\email{\{name.surname\}@uniba.it}\and
KDDLab - ISTI - CNR, Pisa, Italy\\
\email{\{name.surname\}@isti.cnr.it}
}
\maketitle              
\begin{abstract}
Artificial Intelligence algorithms have now become pervasive in multiple high-stakes domains. However, their internal logic can be obscure to humans.
Explainable Artificial Intelligence aims to design tools and techniques to illustrate the predictions of the so-called black-box algorithms.
The Human-Computer Interaction community has long stressed the need for a more user-centered approach to Explainable AI. This approach can benefit from research in user interface, user experience, and visual analytics. This paper proposes a visual-based method to illustrate rules paired with feature importance.
A user study with 15 participants was conducted comparing our visual method with the original output of the algorithm and textual representation to test its effectiveness with users.

\keywords{ User-centric Explainable AI \and Visual Analytics \and User Interfaces for Explainable AI}
\end{abstract}
\section{Introduction}
Artificial Intelligence (AI) and Machine Learning (ML) decision-making systems are widely used in high-stakes domains such as healthcare, justice, and finance. Their usefulness in solving increasingly complex tasks comes at a cost: the internal logic behind the model is often unintelligible to humans. Following the General Data Protection Regulations (GDPR) by the European Union, which establishes a right to an explanation for a user affected by an automated decision-making system~\cite{GDPR}, there has been an emergence of explainable artificial intelligence (XAI) techniques in recent years~\cite{abdul2018trends, liao2020questioning, guidotti2018survey}. These techniques seek to make AI and ML models interpretable by humans. However, several studies ~\cite{miller2019explanation,cheng2019explaining,lipton2018mythos, abdul2018trends,liao2021human} have pointed out that most of the work in XAI is built upon researchers’ belief of what a ``good'' explanation  is~\cite{cheng2019explaining, miller2019explanation}, framing XAI mainly as an algorithmic problem and not focusing enough on the user's point of view~\cite{millerinmates2017}. 
The Human-Computer Interaction (HCI) community has recently directed its attention toward the intersection of Artificial Intelligence (AI) and Explainable AI (XAI), incorporating theories and concepts from the rich research field of HCI and reframing XAI as a design problem\cite{liao2021human}. In the present study, we introduce a novel and innovative approach named FIPER (Feature Importance Plot for Explanatory Rules) that leverages the visualization of explanations through the fusion of rules and feature importance. While rules play a central role in our investigation, we augment their significance by incorporating feature importance, which serves to provide valuable context regarding the role and relevance of a specific feature upon which a rule's predicate has been generated. By integrating these two components, we anticipate a significant enhancement in the user experience, facilitating a more comprehensive and intuitive understanding of the underlying prediction algorithm.


The central role in this study is played by rules; however, they are supported by Feature Importance(FI). FI provides context for the significance and relevance of a feature for which a rule predicate has been formulated. This integration is expected to enhance both the user experience and the user's conceptual understanding of the prediction algorithm.
\FIPER{} exploits interactivity by design to manage datasets with a high number of features. The user can filter the attributes to focus only on those that are predicates of the rule. In this way, the cognitive workload is lowered. 
\FIPER{} is designed around the data scientist/developer enabling the user to verify hypotheses and to assess that the AI model works accordingly to the expected behavior~\cite{preece2018stakeholders}.
A user study was performed to investigate if \FIPER{} effectively supports the users in their work. ~\cite{abdul2018trends, mucha2021interfaces}.

\section{Related Work}
This work contributes to the field of human-centered explainable AI, which seeks to bring techniques and methodologies from HCI into the design of explanations~\cite{liao2021human}.  Our primary objective is to leverage a visual representation of the explanation to facilitate and enhance human interpretation.
In 2018 the Defense Advanced Research Projects Agency (DARPA) ``explainable AI initiative'' framed the explainable AI process as a three-stage approach, distinguishing between the explainable model, the explanation user interface, and the psychological requirements crucial for their design. By differentiating between the model responsible for generating explanations for machine learning algorithms and the means employed to effectively communicate these explanations to the user, this framework provides a comprehensive understanding of the multifaceted nature of explainable AI.

By aligning our research with these established frameworks and principles, we aim to contribute to advancing the field by proposing a novel visual-based approach that caters to the psychological requirements of users. By focusing on the design of an intuitive and visually appealing explanation user interface, we aspire to bridge the gap between complex machine learning models and human comprehension, thereby enabling users to gain meaningful insights and a deeper understanding of the underlying AI algorithms~\cite{gunning2019darpa, chromik2021human}. 

Chromik \textit{et al.} define an explanation user interface (XUI) \emph{``as the sum of outputs of an XAI system that the user can directly interact with. An XUI may tap into the ML model or may use one or more explanation-generating algorithms to provide relevant insights for a particular audience''}~\cite{chromik2021human}. The separation between the explainable algorithm and how the explanation is presented to the users has also been pointed out by~\cite{danilevsky2020survey}: the authors differentiate between explanation techniques and explanation visualizations. The first involves the generation of \emph{rough explanations}, usually propounded by AI researchers, while the latter concerns how these rough explanations are presented to users.
Text can be used to convey a simple form of explanation, while
the conjunctive use of text and visual cues can enhance how explanations are delivered to the users~\cite{ehsan2019automated}. However, visualization is better suited to communicate complex concepts~\cite{andrienko2022visual}. ~\cite{wang2019designing} cite basic charts for raw data, as well as tornado diagrams for list attribution, and saliency heatmaps for image-based models.
It has been proven by~\cite{mucha2021interfaces} that the way an explanation is displayed has an effect on how the user makes decisions. In~\cite{yang2020visual}, the authors investigate how different visual displays of example-based explanation affect the user appropriate trust of the ML classification. However, as reported by ~\cite{mucha2021interfaces}, the design and testing of different visualizations for Explainable AI are still under-studied. 
The pure text has been used to show rules~\cite{ming2018rulematrix,freitas2014comprehensible}; however, some visual examples  can be found in ~\cite{mucha2021interfaces, ming2018rulematrix,andrienko2022visual}.
Another key point for explanations is the implementation of interactivity: although advocated in several studies, its integration within explanations is still limited~\cite{abdul2018trends,chromik2020taxonomy,madumal2019grounded}
\section{Visual Explanation for Rules and Feature Importance}
\subsection{XAI methods}

 
We address the problem of representing explanations based on rules in a visual format to enable the user to investigate the relationship of the input with the outcome of the decision system. Accordingly to~\cite{guidotti2018local}, a rule can be formally defined as a statement like $p \rightarrow y$, where the \textit{consequence} $y$ is the output of the black-box and the \textit{premise} $p$ is a conjunction of split conditions on the observed features, where each condition can be represented as a predicate of the form $a_i \in [v_{i,l}, v_{i,h}]$, where $a_i$ is one of the features of the data and $v_{i,l}$ and $v_{i,h}$ are respectively the lower and upper bounds for the domain of $a_i$ where the predicate is valid. For categorical data types the predicate has the form $a_i \in \{v_l, v_j,\dots,v_k\}$, where each $v_i$ is a value of $a_i$. An instance $x$ is covered by a rule $r$ if all the predicates of the premise of $r$ are satisfied by $x$.

Automatic scripts and programs can efficiently manage rules to enable support for reasoning and exploration. However, this formal representation may present a high cognitive load for the user. Moreover, rule predicates do not provide an explicit ranking of the features of the data. Explanation methods based on Feature Importance provide a ranking of each feature based on the relevance of the feature in the final decision. Given an instance $x$, a FI method returns an ordered sequence of pairs $[(a_j, w_1), (a_l, w_2), \dots, (a_k, w_m)]$, where each feature $a_i$ is associated with a weight $w_i$ that represents the relevance of $a_i$ in the decision. For local explanation, the reference to a feature $a_i$ intentionally means the actual value observed for $a_i$ in the current instance $x$. The explanations based on FI are lightweight from a cognitive point of view, although they provide less information than those based on rules. 

We propose a visual interface that combines the strong points of both groups of explanation strategies. In particular, we exploit FI to enforce a ranking on the visualized features to guarantee that the most relevant are the firsts shown to the user. We introduce a visual encoding to represent the intervals yielded by the rule predicates to easily catch the relationship of each interval with the global distribution of the data. 
Without loss in generality, we identified two methods for both families of explanation strategies: LORE ~\cite{guidotti2018local} and  SHAP~\cite{lundberg2017unified}.
We opted for LORE due to prior experience of its adoption in previous case studies. Nonetheless, the visualization is versatile enough to accommodate other rule-generating algorithms like Anchor~\cite{Anchors2018}, by adopting a translation interface to match the input schema of our tool. For calculating Feature Importance, we employed SHAP, a widely recognized standard in the field. Alternately, other algorithms such as LIME~\cite{ribeiro2016should}, which also generate Feature Importance, can be employed.


\begin{figure}[ht!]
\includegraphics[width=1.1\linewidth]{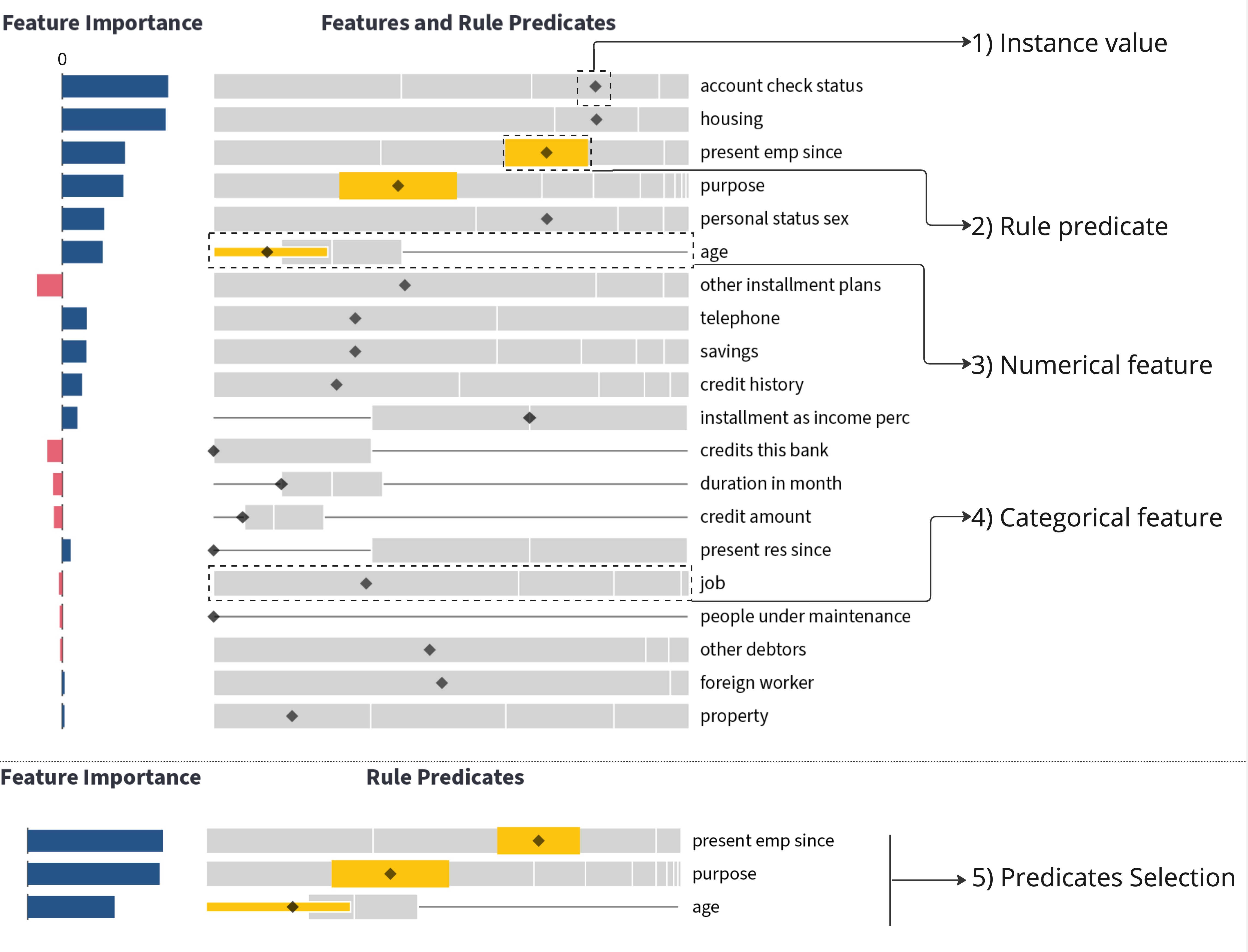}
    \caption{\FIPER{} visualization of one instance of the \textit{German Credit Risk} dataset. \textit{(Top)} Attributes are sorted by the absolute value of FI. Categorical attributes are represented as stacked absolute bar charts. Numerical values are represented as box plots. The interval contained in the predicates of the rule are highlighted in yellow. \textit{(Bottom)} Filtered view of the visualization, showing only the attributes referred in the rule premise}
    \label{fig:fiper_example}
\end{figure}
\subsection{Visualization}
Our proposal organizes the explanation's visual space to combine the information yielded by the FI and the rule-based methods. Figure~\ref{fig:fiper_example} shows a visualization of an instance extracted from the \textit{German Credit Risk} dataset from UCI~\cite{Dua:2019}, a dataset widely used for educational purposes. The visualization comprises two panels: on the left, the weights of the FI methods are reported and sorted accordingly to their absolute values; on the right, the rule predicates are visualized following the order of the first diagram. The FI panel represents the weights by color coding the corresponding feature's positive (blue) or negative (magenta) contribution. The visualization is designed using a color-blind-friendly palette. The rule predicates panel visualizes all the features with a specific chart aligned with the elements in the FI panel. We use two different types of charts based on the type of each feature. For \textit{categorical data types}, we adopt a stacked bar chart to show the part-of-the-whole relationship of each possible value. With this representation, the user can catch the internal distribution of the values. A diamond point is located in the center of the value observed for the attribute in $x$.  For \textit{numerical data types}, we use a box plot chart that shows a compact visualization of the data distribution: min, max, first quartile, third quartile, and median. The observed value for $x$ is represented by a diamond point located within the scale of the box plot. For those attributes for which exists a predicate $p$ in the rule $r$ we add a second layer to highlight the intervals of the rule. The interval visualization changes accordingly to the data type. For categorical data, the intervals contained in the rule premise are highlighted in yellow. For numerical data, a yellow bar represents the extent of the predicate values.

\begin{figure}[htb]
     \centering
         \includegraphics[width=0.9\textwidth]{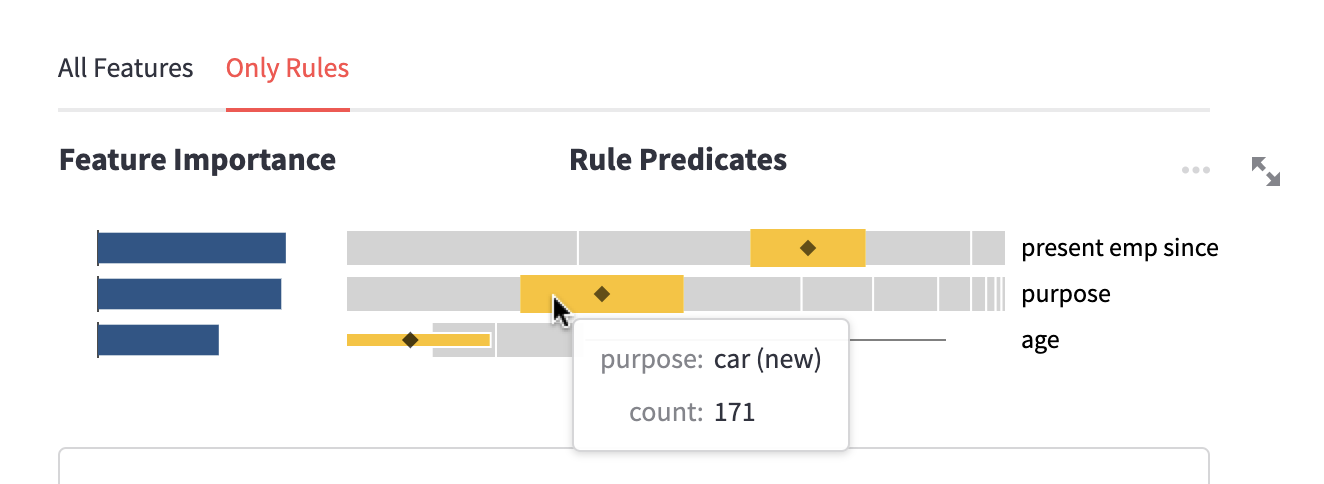}
         \includegraphics[width=0.9\textwidth]{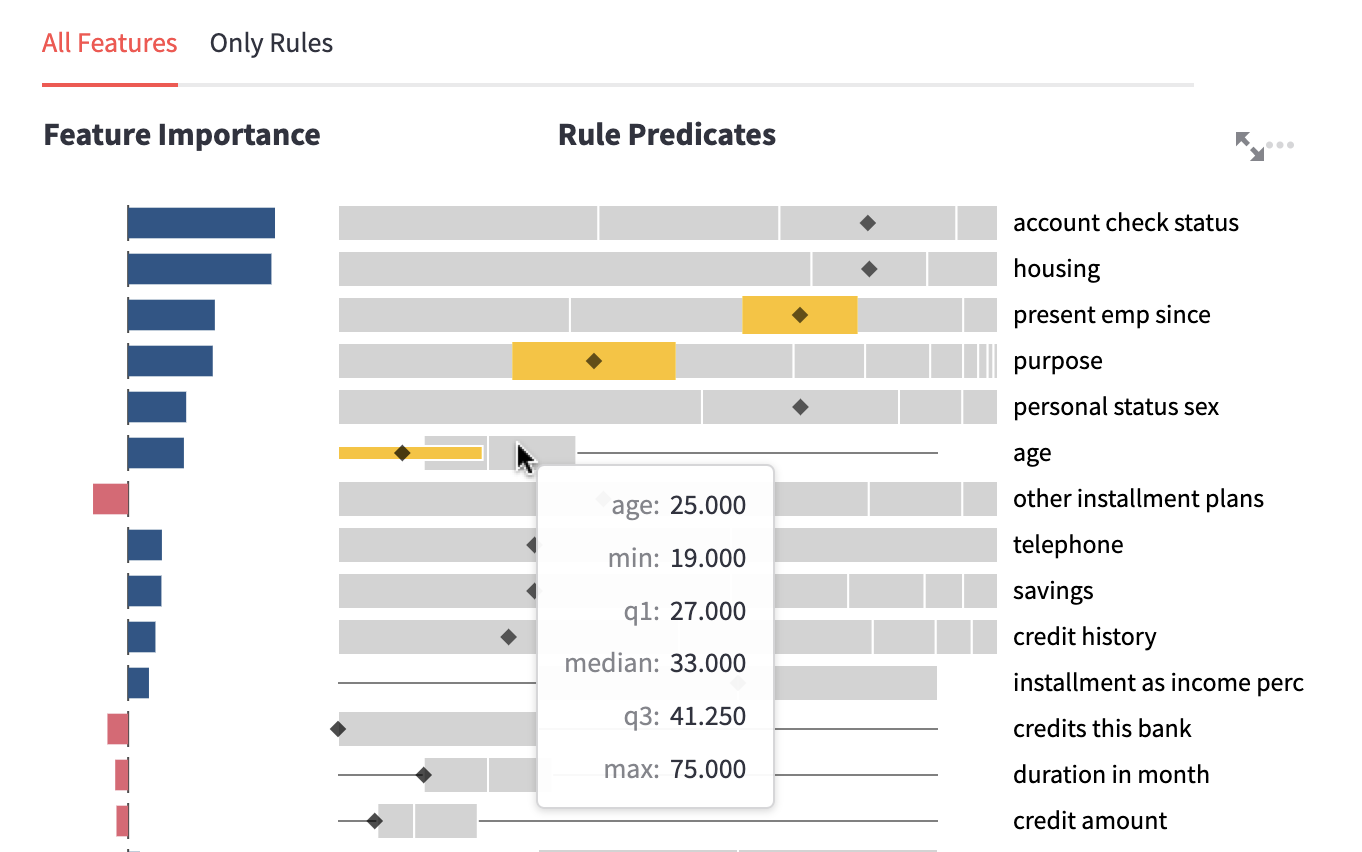}
        \caption{Finer details of a specific feature, selected by hovering the mouse on the corresponding row. (Top) Tooltip for a categorical data type, where the feature's actual value is shown with its class's cardinality. (Bottom) Tooltip for a numerical data type, where statistical central values are shown: min, max, median, Q1, and Q3.}
        \label{fig:fiper_hover}
\end{figure}

The example in Figure~\ref{fig:fiper_example} shows a rule for an instance of the dataset that is classified as \textit{Bad Credit Risk}. The attributes are sorted by the absolute values of weights of FI. For instance, the attributes \textit{account check status} and \textit{housing} (both categorical) are the most relevant for FI, even if they are not mentioned by the rule associated with the prediction. 
The predicates of the rule refer to three attributes: \textit{present employed since}, \textit{purpose}, and \textit{age}. The interval for the predicate for age is relevant since it covers the lower part of the distribution. The diamond shows that the associated value is below the first quartile. 
As suggested by ~\cite{abdul2018trends, miller2019explanation,chromik2021human}, two forms of interactivity are implemented:
\begin{itemize}
    \item To focus  the user's attention only on the predicates, it is possible to dynamically restrict the view only to those attributes mentioned within the rule. This interaction follows the principles of giving users easy access to relevant and important information~\cite{schaffer2015getting,lim2010toolkit}.
    The lower part of Figure~\ref{fig:fiper_example}\textit{(Bottom)} shows the restricted version.
    \item For each attribute, the user may get access to a finer level of details of the corresponding distribution by hovering the pointer over the visualization. Figure~\ref{fig:fiper_hover}  shows two different styles of tooltips for two different data types.  For the categorical data type (Top), we show the selected value and its cardinality. For numerical data type, we show a set of representative values (min, max, Q1, Q3, median) and the value of the corresponding feature.
\end{itemize}

\subsection{Other explanation modalities.}
\begin{figure}[ht]

\begin{subfigure}{0.65\textwidth}
\includegraphics[width=1\linewidth ]{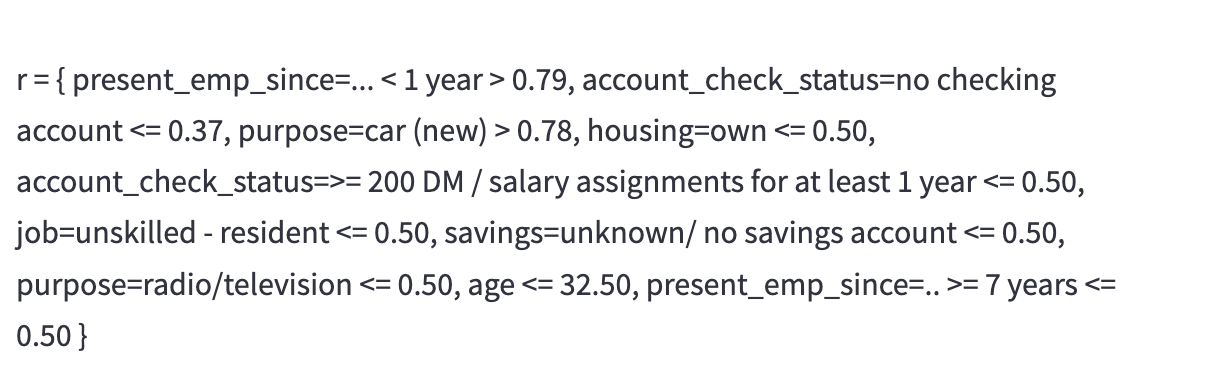} 
\caption{LORE Output}
\label{fig:subim1}
\end{subfigure}
\begin{subfigure}{0.34\textwidth}
\includegraphics[width=\linewidth]{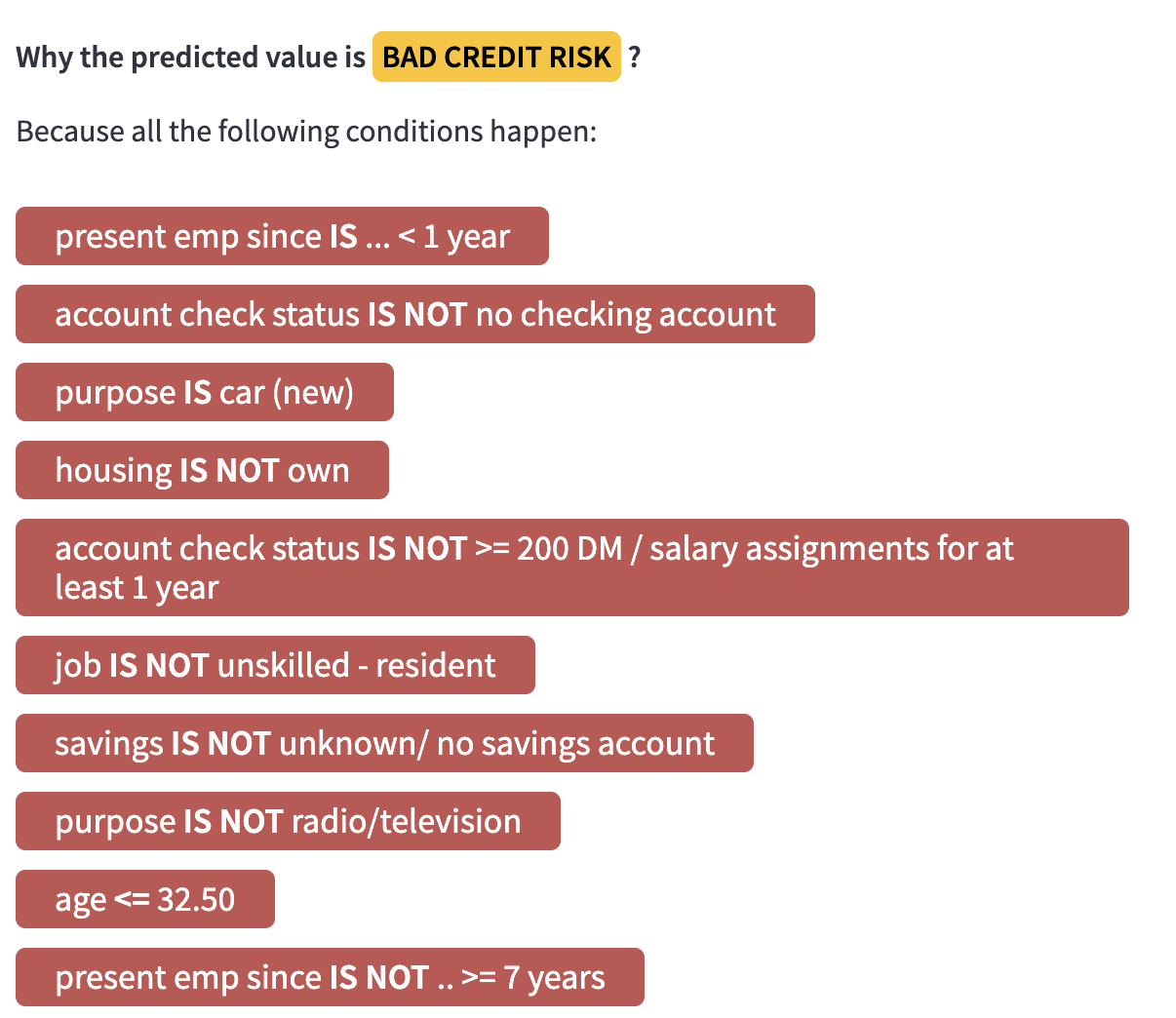}
\caption{XAI Library}
\label{fig:subim2}
\end{subfigure}

\caption{The same instance of Figure\ref{fig:fiper_example} visualized as LORE output and XAI library visualization}
\label{fig:lore_XAI}
\end{figure}

We compared \FIPER{} with two other interfaces, presented in Figure~\ref{fig:lore_XAI}. LORE output is the raw output of the algorithm as text. The XAI library visualization is the implementation already available within the XAI Library\footnote{\texttt{https://pypi.org/project/XAI-Library/}}. The latter enhances the rule's content by improving each predicate's readability with a sequence of graphical blocks.

\section{User Study}
A user study was performed to better understand the value of \FIPER{} in providing clear and understandable explanations. More specifically, three different explanation modalities were compared (i.e., Lore simple output, XAI library output, and \FIPER{}) to answer the following research questions: 
\begin{itemize}
    \item \textbf{RQ1}. Can the explanation modalities support data scientists in understanding the AI model? 
    \item \textbf{RQ2}. What is the difference among the explanation modalities regarding data scientist satisfaction?
\end{itemize}
\subsection{Participants and design}
A total of 15 students in the second year of the Master’s degree in Data Science (7 females) participated in the study. Their mean age was 25 years (SD = 2.7, min = 22, max = 31). Out of 15 students, 12 declared their knowledge of the dataset as "very low" and "low", while 3 stated that they had good knowledge of it.
Considering the number of participants, a within-subject design was performed ~\cite{preece2015interaction}, with the explanation modality as an independent variable and three within-subject factors, i.e., Lore simple output, XAI library output, and \FIPER{}. The participants worked individually with the three modalities and provided their opinion. 

\subsection{The experimental tasks} 
The participants were asked to carry out a sequence of 3 instances of increasing difficulty, defined by the number of predicates in the rule. For each instance, the participant had to answer three experimental questions. The first one asked them to indicate which features are present within the rule. The second question requested them to identify (if any) rule predicates that are insignificant to the prediction. Finally, they were asked to identify the most relevant predicate of the rule to determine the prediction. Because each of the 15 participants answered the above 3 questions for each of the 3 instances on 3 modalities, the total number of answers collected was 405 ($15 \times 3 \times 3 \times 3$). To avoid possible unfair effects of learning from the first task (i.e. order effect)~\cite{sharp2019interaction}, the questions and the explanation modality order were counterbalanced across the participants, according to a Latin Square design.

\subsection{Procedure} 
The study occurred in a quiet university room where the students attended their classes. Three researchers were involved, who intervened just if technical problems emerged. The study lasted one hour and a half, starting from the presentation of the study goal to the participants, including the interaction with the three visualization modalities to answer the experimental questions, until the completion of questionnaires administered before and after the interaction to collect data about the participants and their opinion on visualizations. 
All participants followed the same procedure. First, they were introduced to the study purpose and what they had to do. Participants were asked to sign an informed consent as our university's ethics committee requires for the user study. All participants provided consent. Then, the participants were invited to the study via a link to a web platform that allowed them to answer the experimental questions using the three modalities. Once the participant clicked on the link, a page providing an overview of the study and its goals appeared. At this stage, the platform requested participants to fill in a questionnaire to collect their demographic data, and their familiarity with the domain on a scale from 1 to 7 (1 - being not familiar at all, 7 - being very familiar). The other data were collected anonymously, with no means of identifying individual participants. The platform randomly assigned participants to one of the three explanation modalities. Thus, a training session where they saw the instructions on how to read each explanation was followed by one practice trial.
The actual study session then started. The participant interacted with the first visualization modality to answer the three experimental questions for each instance. Then, the participant completed an online questionnaire including NASA-TLX~\cite{hart2006nasa}. This procedure was the same for all the 3 conditions. However, before repeating the same procedure with the next modality, the participants were invited to relax for 5 minutes. Finally, the platform asked participants to fill in a final questionnaire to express their satisfaction with the modalities they had just used and to vote for the best explanation modality and explain why; the questionnaire included the User Engagement Scale (UES)~\cite{o2018practical} in its short form related to the preferred visualization. At the end of the study, participants were thanked for their participation. A pilot study involving three participants was conducted to check the overall research methodology.

\subsection{Data collection and analysis.}
Quantitative and qualitative data were collected to answer the two research questions. 
To analyze the support (RQ1) provided by the explanation modalities to data scientists, metrics such as the error rate and task execution time were considered.

\subsection{Error rate.} Participants could make two different types of errors while performing the tasks: when asked to list a set of features, they could either enter a feature that was not present (E1) or not enter a feature that was present (E2). To calculate the error rate, we created nine vectors containing the correct answers, three for each instance. The elements of the vectors were equal to the number of features in the train set of the dataset, plus the ``I' don't know'' option. The 9 vectors were compared with those of the responses given by the users to compute the error rate; in Figure ~\ref{fig:my_label}, the error rate of Lore Output is compared to the other two visualizations.
During the task for each instance and visualization, we noted the completion time each participant took to analyze the visualizations and complete the task. We use these time measurements as a proxy for the cognitive workload of each condition.

\subsection{User satisfaction.}The online questionnaire to investigate satisfaction (RQ2) with the explanation modality was composed of two sections. The first section proposed the NASA-TLX questionnaire, used as “Raw TLX” ~\cite{hart2006nasa}. It is a 6-item survey that rates perceived workload using a system through 6 subjective dimensions, i.e., Mental Demand, Physical Demand, Temporal Demand, Performance, Effort and Frustration, rated within a 100-point range with 5-point steps (lower is better). These ratings were combined to calculate the overall NASA-TLX workload index ~\cite{hart1988development}. Specifically, the NASA-TLX was used to assess the workload caused by each modality because the user's workload when using a software tool influences user satisfaction. The second section presented the new UES (User Engagement Scale) short form, derived from the UES long form. It is a 12-item survey used to measure user engagement, a quality characterized by the depth of a user’s investment when interacting with a digital system ~\cite{o2016theoretical,}. It typically results in positive outcomes~\cite{o2018practical}. This tool measures user engagement by summarizing an index that ranges from 0 to 5. It also provides detailed information about four dimensions of user engagement, i.e., Focused Attention (FA), Perceived Usability (PU), Aesthetic Appeal (AE), and Reward (RW). 
The last questionnaire was administered when the participants used all three explanation modalities. It evaluated the participant's satisfaction by asking them to rank the three modalities based on their Utility, Completeness, Understandability, and Helpfulness (from 1 to 3, 1 is the best) and to vote for the best visualization explaining why they preferred a modality over the others.

\begin{figure}[ht!]
    \centering
    \includegraphics[width=\textwidth]{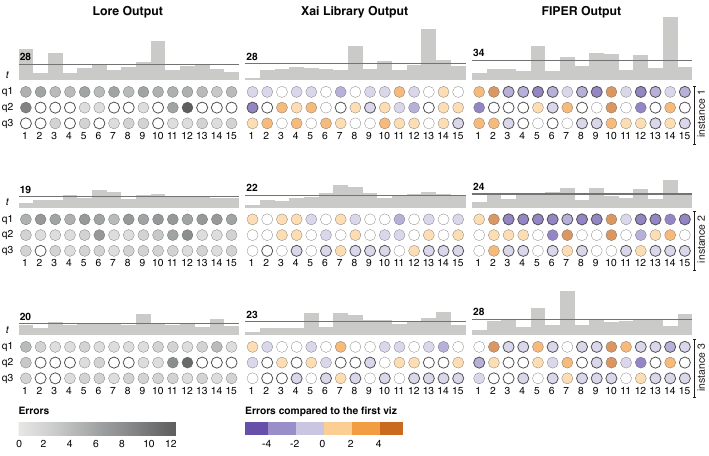}
    \includegraphics[width=\textwidth]{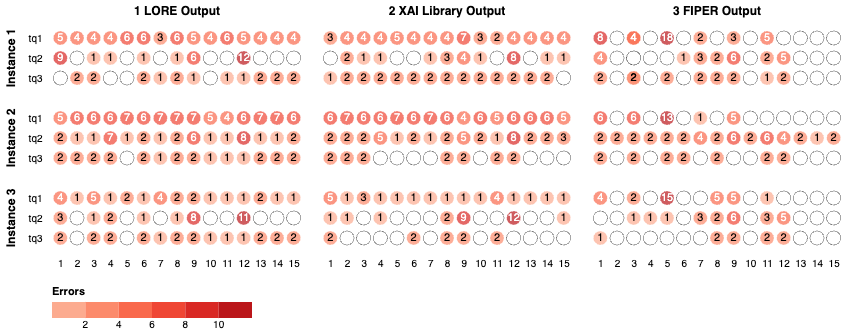}
    \caption{Errors rate and completion times for tasks in each condition.\textit{(Top)} The first column shows the absolute errors of the LORE output. The other two columns show the difference w.r.t. the first column, with divergent color scale to highlight increment or decrement in errors. \textit{(Bottom)} Absolute number of errors for each output and each task}
    \label{fig:delta_errors}
\end{figure}

\subsection{Results}

Figure~\ref{fig:delta_errors}\textit{(Top)} shows all tasks' errors and completion times for all the users. The results are organized by instance (rows) and visualizations (columns). For each cell of this grid, the chart shows each user's actual time to complete the three tasks (bar chart on the top, with a line showing the median completion time) and the number of errors (heatmap on the bottom). We use two distinct colors scales for the heatmaps. The LORE Output visualization (first column) uses grayscale to represent each task's absolute number of errors. The other two visualizations (columns 2 and 3) show the difference from the LORE Output errors using a divergent color scale: purples maps to better performance and oranges to higher errors. The tasks with no errors are denoted with a thicker black stroke. Figure~\ref{fig:delta_errors}\textit{(Bottom)} reports the absolute errors for each task and each condition. In this case, each circle has a color proportional to the number of errors. Each circle reports the actual number for further details. Gray circles denote tasks where there are no errors.
Although considered the easiest, the first instance shown to the user required a higher completion time. This might be because the user had to get acquainted with the different visualization strategies and gain the correct way to read the outputs. Among the three visualizations, \FIPER{} appears to be the most time-consuming but with a significant improvement in the error rate. 
We can conclude from this observation that although \FIPER{} is slightly more time-demanding than the other two visualizations, it performs better on all the tasks, even with the most difficult instances. From the user satisfaction questionnaire, 
\FIPER{} visualization was chosen as the top preferred by 10  participants, 4 preferred the XAI Library visualization, and only one chose Lore simple output. \FIPER{} was considered the most valuable visualization by 13 participants (the other two chose the XAI library). 
Overall the \FIPER{} was considered more understandable by 10 participants.

The XAI library Visualization was appreciated by some of the participants for its conciseness, and it was pointed out that it may be more suited for datasets with a low number of features.
A participant commented that \emph{FIPER is the most easily readable visualization even though XAI Library Viz might be more immediate for certain questions}. This follows what was stated by ~\cite{kulesza2013too} about the completeness of explanations which is positively correlated with improved mental models and does not impair user experience or task.


\section{Conclusions and future work}



This research paper introduces a novel visual-based approach for rule representation. It evaluates its effectiveness compared to two existing textual-based approaches within the context of data science education. By integrating the rules with feature importance, we aim to enhance the user experience and promote a better understanding of the underlying prediction algorithm.

The preliminary findings of our study indicate that the visual-based approach demonstrates superior suitability for datasets containing a high number of features. Conversely, qualitative feedback suggests that the XAI library visualization may be more suitable for datasets with fewer attributes. It is worth noting that our initial testing of \FIPER{} was conducted with data scientists; however, we intend to extend its application and customization to various scenarios, specifically targeting experts in different domains. Furthermore, \FIPER{} offers users a certain degree of interactivity, allowing them to engage with the explanations provided.

This work is not free of limitations. Feature importance is used to sort predicates, even if it can contrast with the rule-based approach. However, it's worth noting that rule-based methods don't consistently provide a logical order when displaying rules and the order of presentation of the predicates may not be aligned with their relevance. Thus, our approach makes a design choice to exploit FI ranking in the visualization. 
The user retains the option to decide whether to apply this sorting, for example prioritizing the visualization of those features that are present in one of the predicates of the rule. A future development comprises the possibility for the user to select the FI method among a given set, allowing the users to confront different FI algorithms.

\FIPER{} enables us to assess an instance's ranking within the distributions of individual features. We are enhancing the interface to permit instance editing, granting users the capacity to modify specific feature values and delve into the black box's response.
We are also actively developing a \FIPER{} version that visualizes counter rules, which are logical predicates based on features that result in an alternate classification of the chosen instance.

Eventually, for specific explainers, like LORE in our instance, a synthetic neighborhood is established around the instance to build the explanation. We are working to add a layer of presentation of the distribution of feature values within this neighborhood.


\section{Acknowledgements}
This work has been supported by the European Community Horizon~2020 programme under the funding scheme ERC-2018-ADG G.A. 834756 \textit{XAI: Science and technology for the eXplanation of AI decision making}, by the European Union's Horizon Europe Programme under the CREXDATA project, grant agreement no. 101092749, by the Next Generation EU: NRRP Initiative, Mission 4, Component 2, Investment 1.3, PE0000013 – “Future Artificial Intelligence Research – FAIR” - CUP: H97G22000210007 and  ``SoBigData.it - Strengthening the Italian RI for Social Mining and Big Data Analytics'' - Prot. IR0000013.

\newpage
\printbibliography

\end{document}